\documentclass[sigconf]{acmart}

\usepackage{booktabs}

\AtBeginDocument{%
}

\usepackage{tikz}
\usepackage{float}

\setcopyright{none}

\title{Less Deliberate in Teams: Student LLM Use Across
  Individual and Collaborative Work}

\author{Sehrish Basir Nizamani}
\affiliation{%
  \institution{Virginia Polytechnic Institute and State University}
  \city{Blacksburg}
  \state{Virginia}
  \country{United States}}
\email{sehrishbasir@vt.edu}

\author{Zannah Ziew}
\affiliation{%
  \institution{Virginia Polytechnic Institute and State University}
  \city{Blacksburg}
  \state{Virginia}
  \country{United States}}
\email{zannah24@vt.edu}

\author{Saad Nizamani}
\affiliation{%
  \institution{Virginia Polytechnic Institute and State University}
  \city{Blacksburg}
  \state{Virginia}
  \country{United States}}
\email{saadnizamani@vt.edu}

\author{Khyati Goyal}
\affiliation{%
  \institution{Virginia Polytechnic Institute and State University}
  \city{Blacksburg}
  \state{Virginia}
  \country{United States}}
\email{khyati22@vt.edu}

\begin{abstract}
As large language models (LLMs) become common in computing
courses, we need to understand how the social setting shapes
how students use them. This paper reports findings from a
semester-long study of 96 undergraduate students who completed
six assignments, alternating between individual homework and
team project milestones. We tracked LLM usage, prompting
habits, and how students verified AI-generated output across
all six assignments. LLM usage dropped by 42.7 percentage
points when students moved from individual work to their first
team milestone, then partly recovered in later team tasks.
Students also wrote fewer prompts, used fewer deliberate
prompting strategies, and checked LLM output less
carefully. The share of students who ran tests on AI-generated
code fell by 19.4 percentage points during team assignments
and never fully rebounded. A within-student analysis found
that 18.9\% of students who consistently used LLMs on their
own stopped using them entirely in teams, while only 3.2\%
went the other direction. These results suggest that
collaborative context is associated with reduced deliberate
LLM engagement beyond what task type alone can explain. The
moment students form teams appears to be a critical turning
point that may benefit from more explicit instructional
support.
\end{abstract}

\keywords{large language models, generative AI, collaborative
  learning, team-based assignments, AI literacy, verification
  behavior, computing education, within-subject study,
  assignment design, socially situated learning}

\begin{document}
\maketitle

\section{Introduction}

Large language models (LLMs) are rapidly becoming embedded in
computing education, shaping how students write code, debug,
and reason about problems~\cite{zhu2026, nieto2026}. Prior
work documents widespread adoption of LLMs among computing
students, though with varying levels of trust and
skepticism~\cite{cambaz2024, raihan2025}. Meaningful learning
depends not on use alone, but on how deliberately students
engage with these tools, through prompting, reflection, and
verification~\cite{poitras2024, zhu2026, nieto2026}.

Despite growing interest in LLM use, most studies examine
students in isolation, focusing on individual assignments or
individual outcomes~\cite{raihan2025, pereira2025}. However,
learning in computing contexts is fundamentally shaped by
social interaction: peer networks, collaborative discourse,
and shared tools are central to how students develop
programming knowledge~\cite{stahl2006, hou2025}. Students
work across both individual and team-based assignments, yet
social context fundamentally reshapes how individuals engage
with tools and distribute effort~\cite{stahl2006}. Little is
known about how this transition affects students'
interactions with LLMs.

In this paper, we examine how students' LLM behavior changes
as they move between individual and team-based assignments. We
report findings from a semester-long, within-subject study of
96 undergraduate students completing alternating individual
homework and team project milestones. Across six assignments,
we tracked LLM usage, prompting behavior, and verification
practices.

Our findings reveal a consistent pattern not previously
examined through within-subject comparison across individual
and team contexts. LLM usage drops sharply at the onset of
team work before partially recovering in later team
assignments. However,
this recovery is incomplete: even when task demands are
comparable, students in teams use LLMs less frequently and
engage with them less deliberately. In particular, students
write fewer prompts, employ fewer prompting strategies, and
verify outputs less rigorously. The proportion of students
running tests on AI-generated code decreases substantially in
team settings and does not fully rebound.

Taken together, these results suggest that collaborative
context does not simply reduce LLM use---it reshapes it. The
transition from individual to team work emerges as a critical
point at which students disengage from the deliberate practices
that support effective AI-assisted learning.

To structure this investigation, we address the following
research questions:
\begin{enumerate}
  \item[\textbf{RQ1.}] How does self-reported LLM usage rate
    and role framing change when students move from individual
    homework assignments to team-based project milestones?
  \item[\textbf{RQ2.}] How do prompting frequency and
    prompting technique shift across individual and
    collaborative assignment contexts?
  \item[\textbf{RQ3.}] How do output verification behaviors
    change when students transition from individual to
    team-based work?
  \item[\textbf{RQ4.}] At the individual level, what patterns
    of behavioral transition characterize students who change
    their LLM usage when moving between individual and
    collaborative contexts?
\end{enumerate}

\section{Background and Related Work}

\textbf{LLM use in student programming.}
LLMs are rapidly becoming embedded in computing education,
with students routinely using them for code generation,
debugging, explanation, and iterative
refinement~\cite{cambaz2024, raihan2025, pereira2025}. Across
undergraduate contexts, these systems increasingly function as
non-judgmental on-demand programming assistants, supporting
both problem solving and feedback at
scale~\cite{cambaz2024}. Yet adoption rates alone are a
deceptively shallow metric. What matters pedagogically is not
whether students use LLMs, but how deliberately they engage
with them: through iterative prompting, critical evaluation,
and careful verification of generated
outputs~\cite{poitras2024, zhu2026, nieto2026}. These
practices are context-sensitive and cannot be assumed to remain
stable as the conditions of work change.

\textbf{Social context and AI use.}
Students' use of generative AI is not purely individual but
shaped by the social context in which it occurs. Social
influence including peer norms significantly predicts students'
intention to adopt generative AI tools, shaping not only which
tools they choose but also their willingness and confidence to
engage with them~\cite{korchak2025}. Disclosure and
non-disclosure of AI use often reflect alignment with perceived
peer norms, with concerns about judgment, legitimacy, and
academic integrity leading students to suppress or conceal AI
use in socially visible contexts~\cite{qu2026, adnin2025}.

These dynamics become particularly complex in team-based work.
Kharrufa et al.\ found that students using GenAI in software
engineering team projects encountered specific social
challenges: team members who generated AI code without
understanding it created difficulties for others, and the lack
of communication about AI use within teams led to subpar
contributions and wasted effort~\cite{kharrufa2026}. Westerman
and Carruthers similarly observed that students working in
teams held widely varying attitudes toward LLM use, creating
hidden tensions that went unaddressed without explicit team
negotiation~\cite{westerman2026}. These studies establish that
collaborative contexts introduce social dynamics around AI use
that individual-focused research does not capture. Yet neither
study compared LLM behavior across individual and collaborative
assignment contexts, nor examined how the same students changed
their AI practices when moving between the two.

\textbf{Task structure and LLM interaction.}
Integrating LLMs reliably into workflows often requires
constraining their outputs to ensure consistency and
predictability~\cite{liu2024}. Structured workflows typically
require users to validate outputs against explicit
expectations, while open-ended workflows allow more flexible
and exploratory interaction~\cite{liu2024}. Empirical evidence
reflects this distinction: LLMs achieve high accuracy when
extracting explicit or well-defined information, but perform
substantially worse on tasks requiring subjective
interpretation or nuanced reasoning~\cite{mahmoudi2025}.

\textbf{Verification of LLM-generated code.}
Across programming and software engineering contexts,
verification traditionally relies on automated testing,
static analysis, and formal
methods~\cite{liu2023evalplus}. Among these, test-based
evaluation remains the most widely used approach, where code
is assessed based on its ability to pass predefined or
automatically generated test cases~\cite{liu2023evalplus}.
However, the increasing use of LLM-generated code introduces
new challenges. Generated outputs may appear correct while
containing subtle logical errors, and the process of producing
code is often opaque to the user. While recent work explores
alternative paradigms such as LLM-assisted semantic
evaluation~\cite{tong2024codejudge}, existing approaches
largely extend traditional verification methods rather than
rethinking them for AI-assisted programming. In educational
contexts, verification is not only a correctness check but a
critical learning practice~\cite{poitras2024}. Yet it remains
unclear how students engage in verification when working with
LLM-generated code, and whether they apply these practices
consistently across different social settings.

\textbf{The gap this study addresses.}
Prior work has established that LLM use is widespread,
behaviorally variable, and sensitive to social context. Recent
work has begun to examine LLM use within team projects and
the social tensions it introduces~\cite{kharrufa2026, westerman2026}. However, no
study has compared LLM behavior \emph{across} individual and
collaborative assignment contexts within the same students,
examining not only whether usage changes, but how prompting
strategies and verification practices shift when the social
structure of work changes. This gap is consequential: how
students manage AI practices across individual and team-based
assignments has direct implications for learning and equitable
participation, yet no study has examined it through a
within-subject longitudinal design.

\section{Study Design}

\subsection{Course Context and Participants}

The study was conducted across two sections of an upper-division
undergraduate course in data analytics and visualization at a
large public research university in the United States. Both
sections were taught concurrently during the same semester by
two instructors, both of whom are co-authors of this paper.
A total of 96 students across both sections consented to
participate and are included in all analyses reported here,
representing a consent rate of approximately 69\% of the
140 enrolled students (Section~1: $n = 39$;
Section~2: $n = 101$). Both sections followed an identical
curriculum, including the same assignments, rubrics, grading
criteria, and guidance on LLM use. Assignment schedules and
survey timing were synchronized across sections to minimize
potential timing-related effects on observed behavior.

The curriculum was structured to enable a natural
\emph{within-subject} comparison by interleaving individual
and team-based assignments throughout the semester. Rather
than separating individual and collaborative work into
distinct phases, students engaged in both modes in close
temporal proximity. The sequence of assignments included
three individual homework tasks (HW3, HW4, HW5) and three
team project milestones (TP1, TP2, TP3), with team formation
occurring prior to TP1 and remaining fixed
thereafter.

Although individual and team assignments were interleaved
across the semester, they were not always strictly sequential.
One pair of assignments, HW4 and TP1, shared the same
deadline, reflecting the parallel demands typical of project-based
course workflows. Table~\ref{tab:assignments} summarizes the
assignments and their content, and Table~\ref{tab:timeline}
shows their distribution across the semester.

\begin{table}[ht]
  \caption{Overview of the six assignments included in the
    study, alternating between individual homework (HW) and
    team project milestones (TP) across the semester.}
  \label{tab:assignments}
  \begin{tabular}{llp{4.5cm}}
    \toprule
    Assignment & Type & Description \\
    \midrule
    HW3 & Individual &
      Election analysis: data joins, correlation, and
      regression on 2020 county-level results and census
      demographics. \\
    \addlinespace
    TP1 & Team &
      Project initiation: topic selection, research questions,
      and data source identification. \\
    \addlinespace
    HW4 & Individual &
      Clustering and dimension reduction: k-means and PCA
      on US census demographics. \\
    \addlinespace
    TP2 & Team &
      First iteration: data collection, cleaning, exploratory
      analysis, and initial findings. \\
    \addlinespace
    HW5 & Individual &
      Text and sentiment analysis: bag-of-words, sentiment
      scoring, and TFIDF on Yelp reviews. \\
    \addlinespace
    TP3 & Team &
      Final iteration: completed analysis, written report,
      and class presentation. \\
    \bottomrule
  \end{tabular}
\end{table}

\begin{table}[ht]
  \caption{Timeline of the six assignments across the semester,
    alternating between individual homework (HW) and team
    project milestones (TP).}
  \label{tab:timeline}
  \centering
  \begin{minipage}{0.5\linewidth}
    \begin{tabular}{lll}
      \toprule
      Assignment & Type & Deadline \\
      \midrule
      HW3 & Individual & Sep 26 \\
      HW4 & Individual & Oct 24 \\
      TP1 & Team       & Oct 24 \\
      TP2 & Team       & Nov 7  \\
      HW5 & Individual & Nov 14 \\
      TP3 & Team       & Dec 3  \\
      \bottomrule
    \end{tabular}
  \end{minipage}
\end{table}

This interleaved design has an important methodological
implication. Because individual and team assignments were
distributed across the same period of the semester, observed
differences in LLM behavior between HW and TP assignments are
less likely to be explained by temporal factors such as
increasing course difficulty or accumulating fatigue. Rather
than being separated into distinct phases, individual and
collaborative work occurred in close temporal proximity,
allowing for meaningful within-student comparisons.

The primary systematic difference between assignments was the
social structure of the work, whether students were working
alone or as part of a team, though task type, particularly
whether an assignment required coding, also varied across
milestones and is considered alongside social context in the
interpretation of results. This design supports a
within-student analysis that foregrounds social context as a
key variable of interest.

The assignment numbering reflects the broader course structure:
HW3, HW4, and HW5 were the third, fourth, and fifth
assignments in the course, respectively. Data for this study
were collected through post-task surveys administered after
each of these assignments, capturing students' self-reported
LLM use, prompting behavior, and verification practices.
Earlier assignments in the course were not included, as the
survey instrument had not yet been introduced.

Teams worked in a mix of synchronous and asynchronous modes
depending on the milestone and team preference. Students were
not required to meet at scheduled times; coordination
strategies were left to each team's discretion, reflecting
typical project-based course practice.

\paragraph{Researcher positionality and bias mitigation.}
Because members of the research team also served as course
instructors, we took deliberate steps to mitigate potential
influence on participant responses. Survey participation was
voluntary and had no impact on students' grades, and all
responses were de-identified prior to analysis. To further
reduce potential bias, all analyses were conducted by two
researchers who were not affiliated with either course section.
Additionally, students completed surveys individually and
independently, with no visibility into their teammates'
responses.

\subsection{Data Collection}

Data were collected via brief post-task surveys administered
immediately after each of the six assignments, yielding up to
six observation points per student across the semester. The
role categorization used in this study (Assistant, Peer,
Expert, and Did not use) originated from thematic analysis
of student LLM reflections conducted in prior
work~\cite{dur2026}, which identified these three engagement
patterns: treating the LLM as a starting point for their own
reasoning (Assistant), engaging with it collaboratively
(Peer), and relying on it to complete tasks with minimal
personal input (Expert). The post-task survey instrument
used here follows a similar structure subsequently applied
in~\cite{fie2026}. Each survey captured four variables:

\begin{itemize}
  \item \textbf{LLM usage and role framing}
    (\textit{LLMUsage}): Students selected one of four options
    characterizing their relationship to the LLM during that
    assignment: Assistant (helped me generate ideas or
    suggestions), Peer (collaborated with me in a back-and-forth
    process), Expert (I relied heavily on its answers as
    authoritative), or Did not use.
  \item \textbf{Prompt frequency} (\textit{Prompt}): Students
    reported how many prompts they issued using an ordinal
    range: 0, 1--2, 3--5, or 6 or more.
  \item \textbf{Prompting technique} (\textit{Approach}):
    Students selected all applicable techniques from a
    multi-select list including iterative refinement, single
    direct question, copy-paste code or data for debugging,
    step-by-step prompting, and no prompting technique used.
  \item \textbf{Verification method}
    (\textit{VerificationMethod}): Students selected all
    applicable methods by which they verified LLM outputs
    before use, including manual reasoning, running tests,
    cross-checking with web search or documentation, and
    I did not verify the outputs.
\end{itemize}

Survey participation was voluntary, resulting in 90--96
responses per assignment across the six data collection
points. All statistical analyses were conducted in Python.
Analysis scripts were generated with AI assistance and
verified by the authors prior to use to ensure they
performed as intended.

\section{Results}

We report results across four aspects of student LLM use
corresponding to RQ1--RQ4: how students used and framed LLMs,
how they prompted them, how they verified outputs, and how
these behaviors changed at the individual level. For prompting
technique and verification method, students could select
multiple responses, so percentages may sum to more than
100\%.

Unless otherwise noted, all tests are two-tailed with
$\alpha = .05$. Effect sizes are reported as Cram\'{e}r's $V$
for chi-square tests and Cohen's $d$ for the paired $t$-test,
using conventional benchmarks of 0.1/0.3/0.5
(small/medium/large) for $V$ and 0.2/0.5/0.8 for $d$.
Per-assignment usage rates are computed using the full
consented sample of 96 as the denominator for all
assignments, treating survey non-response as non-use. At TP1
specifically, 92 students submitted a survey response and 4
did not; those 4 are counted as non-users for that
assignment.

\subsection{RQ1: LLM Usage Rates and Role Framing}

Across the three individual homework assignments, average
self-reported LLM usage was stable and high: 82.3\% at HW3,
85.4\% at HW4, and 84.4\% at HW5. This stability is notable
given that assignments alternated with team milestones across
the semester, suggesting that individual assignment context
consistently activated LLM use regardless of temporal
position.

The pattern at team milestones was strikingly different and
non-uniform. Usage collapsed to 39.6\% at TP1, a drop of
42.7 percentage points from HW3, before recovering
substantially to 75.0\% at TP2 and 78.1\% at TP3. This
trajectory, collapse followed by near-recovery, was not
anticipated and constitutes the most theoretically significant
pattern in the dataset. We return to its interpretation in
the Discussion.

The sharp drop at TP1 is partly explained by the nature of
the task itself. Unlike the coding-heavy individual homeworks
and later team milestones, TP1 required no coding, teams
were asked only to identify a topic, define research
questions, and locate potential data sources. This reduced
the natural opportunities for LLM engagement. However, even
TP2 and TP3, which involved substantive coding and data
analysis comparable in technical demand to HW4 and
HW5, showed persistently lower usage rates (75.0\% and
78.1\%) than the individual homework average of 84\%, and
verification rates never recovered to individual assignment
levels. This suggests that task type alone does not fully
account for the observed differences, and that collaborative
context is associated with reduced deliberate LLM engagement
beyond what task structure alone can explain.

Beyond raw usage frequency, the quality of LLM engagement
also shifted across contexts. In individual assignments,
students most commonly framed the LLM as an Assistant
(46.4\%) or Peer (36.0\%), with only 12.9\% reporting non-use
and 4.7\% describing the LLM as an Expert whose answers they
accepted as authoritative. Across team assignments, the
proportion reporting non-use more than doubled to 33.7\%,
while Assistant framing fell to 38.0\% and Peer framing fell
to 26.5\%. A chi-square test confirmed that this shift in
role framing distribution was statistically significant
($\chi^2(3) = 35.57$, $p < .0001$, $V = 0.253$, small
effect).

\begin{figure}[h]
  \centering
  \includegraphics[width=\linewidth]{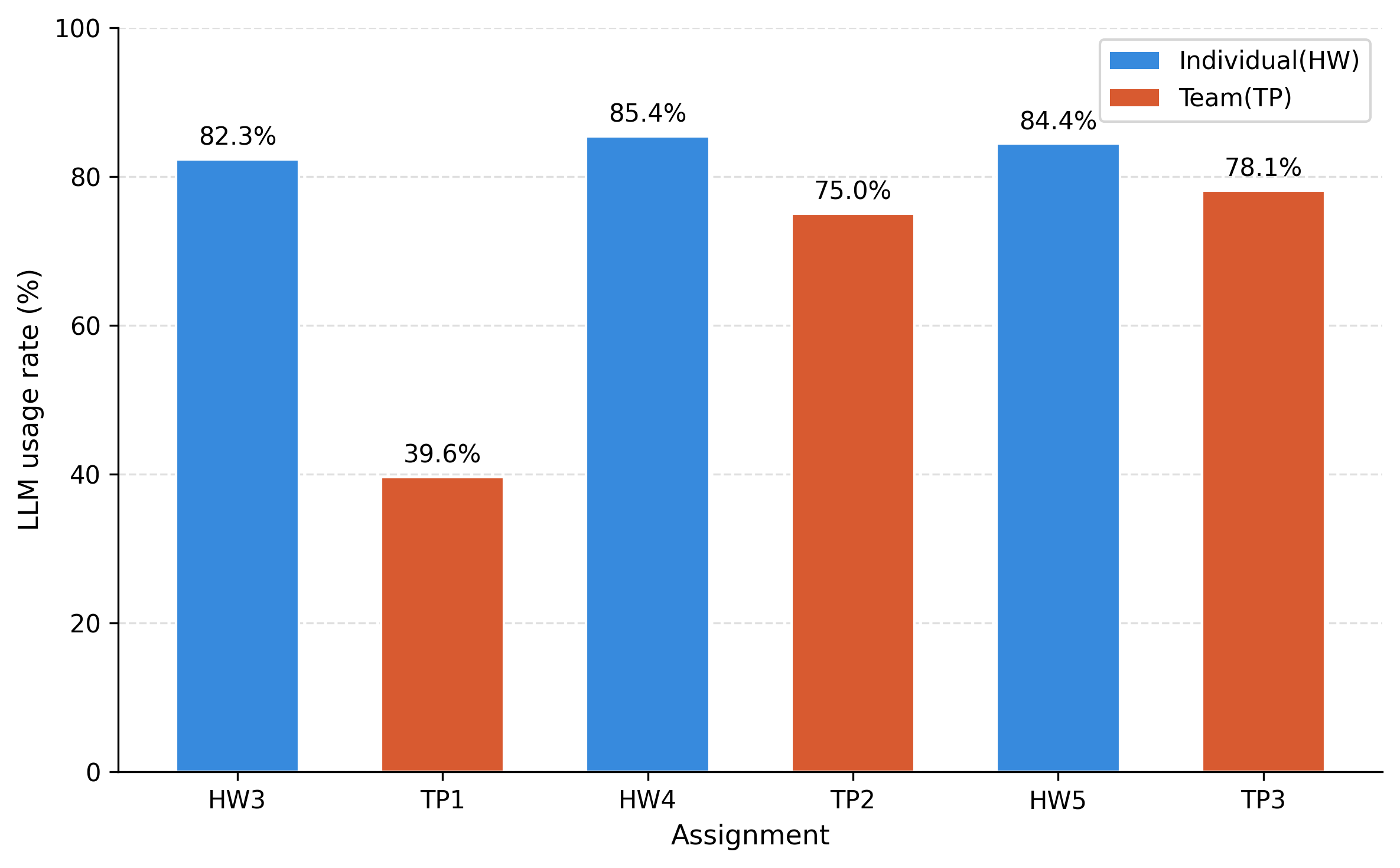}
  \caption{Self-reported LLM usage rates across all six
    assignments in order of completion. Individual homework
    assignments (HW3, HW4, HW5) and team project milestones
    (TP1, TP2, TP3) alternated across the semester. Usage
    collapses sharply at TP1, a non-coding planning task,
    before recovering at TP2 and TP3, which involved
    substantive coding. The persistent gap between TP2/TP3
    and individual homework rates suggests that both task
    type and collaborative context contribute to the
    observed pattern.}
  \Description{A bar chart showing LLM usage percentage for
    each of the six assignments in alternating sequence.
    HW3 is 82.3 percent, TP1 drops to 39.6 percent, HW4
    recovers to 85.4 percent, TP2 is 75.0 percent, HW5 is
    84.4 percent, and TP3 is 78.1 percent. Blue bars
    represent individual homework assignments and orange
    bars represent team project milestones.}
  \label{fig:usage}
\end{figure}

Table~\ref{tab:results} summarizes the key behavioral
measures across both contexts and Figure~\ref{fig:usage}
plots usage rates across all six assignments in sequence.

\begin{table}[h]
  \caption{Summary of key behavioral measures aggregated
    across individual (HW3, HW4, HW5) and team (TP1, TP2,
    TP3) assignments. All differences are statistically
    significant at $p < .0001$.}
  \label{tab:results}
  \begin{tabular}{lccc}
    \toprule
    Measure & Individual & Team & Effect Size \\
    \midrule
    LLM Usage Rate
      & 84\%    & 64.2\%$^{a}$ & --- \\
    ``Did not use'' (role framing)
      & 12.9\%  & 33.7\%       & $V = 0.253$ \\
    Peer or Assistant framing
      & 82.4\%  & 64.5\%       & $V = 0.253$ \\
    Mean prompt count
      & 3.31    & 2.11         & $d = -0.634$ \\
    ``No technique used''
      & 7.3\%   & 29.7\%       & $V = 0.324$ \\
    ``Did not verify''$^{b}$
      & 7.4\%   & 32.6\%       & $V = 0.357$ \\
    Test-based verification$^{b}$
      & 33.2\%  & 13.8\%       & $V = 0.357$ \\
    \bottomrule
  \end{tabular}
  \smallskip

  \noindent\small $^{a}$Team average is elevated by TP2 and
  TP3 recovery; TP1 alone was 39.6\%. See
  Figure~\ref{fig:usage}.

  \noindent\small $^{b}$Verification percentages are computed
  from pooled response counts across all three assignments in
  each context and differ from per-assignment averages cited
  in the text (e.g., manual reasoning: 69.9\% HW average;
  running tests: 54.6\% HW average), which reflect the mean
  of each assignment's respondent base separately.
\end{table}

\subsection{RQ2: Prompting Strategies}

Prompting behavior was measured along two dimensions:
frequency (how many prompts students issued) and technique
(what approach they used).

\paragraph{Prompt frequency.}
Mean prompt counts were substantially higher on individual
assignments (HW mean $= 3.31$) than on team assignments
(TP mean $= 2.11$), a difference of 1.20 prompts per
assignment. A paired $t$-test confirmed this difference was
statistically significant ($t(94) = 5.841$, $p < .0001$,
$d = -0.634$, medium-large effect; $n = 95$, one student
excluded for having no analyzable TP responses). As with usage rates, the
per-assignment breakdown reveals that the aggregate obscures
important variation: prompt counts were lowest at TP1
($M = 0.81$, $SD = 1.30$), nearly one-quarter of the
individual assignment average, before recovering to
$M = 2.59$ at TP2 and $M = 2.96$ at TP3.

\paragraph{Prompting technique.}
The distribution of prompting techniques shifted substantially
between individual and team contexts
($\chi^2(5) = 82.12$, $p < .0001$, $V = 0.324$, medium
effect). On individual assignments, students most commonly
employed iterative refinement (27.9\%), single direct
questions (28.7\%), and copy-paste debugging (17.7\%), with
only 7.3\% reporting no technique. On team assignments, the
proportion reporting no prompting technique used rose to
29.7\%, an increase of 22.4 percentage points, while
copy-paste debugging fell by 11.2 percentage points to
6.5\%. Notably, iterative refinement remained relatively
stable across contexts (27.9\% in HW vs.\ 28.0\% in TP),
suggesting that students who continued to engage deliberately
with LLMs in team settings maintained similar interaction
patterns, while the overall distribution shifted because a
larger proportion disengaged entirely.

\subsection{RQ3: Verification Behaviors}

Output verification showed the most pronounced shift of any
behavioral dimension measured, with the largest effect size
across all measures ($\chi^2(3) = 92.99$, $p < .0001$,
$V = 0.357$, medium effect).

Averaged across the three individual assignments, manual
reasoning was the most common verification method (69.9\%),
followed by running tests (54.6\%), cross-checking with web
search or documentation (28.1\%), and non-verification
(12.2\%). Across team assignments, the proportion reporting
no verification more than quadrupled to 32.6\%, an increase
of 25.2 percentage points. Test-based verification fell from
33.2\% to 13.8\%, a drop of 19.4 percentage points. Manual
reasoning remained nearly unchanged (42.4\% vs.\ 42.2\%).

As with usage and prompting, the TP1 pattern was extreme:
61.4\% of students at TP1 reported no verification at all,
and only 6.0\% ran tests. Both figures recovered meaningfully
by TP2 and TP3, though neither reached individual assignment
levels.

\begin{figure}[H]
  \centering
  \includegraphics[width=\linewidth]{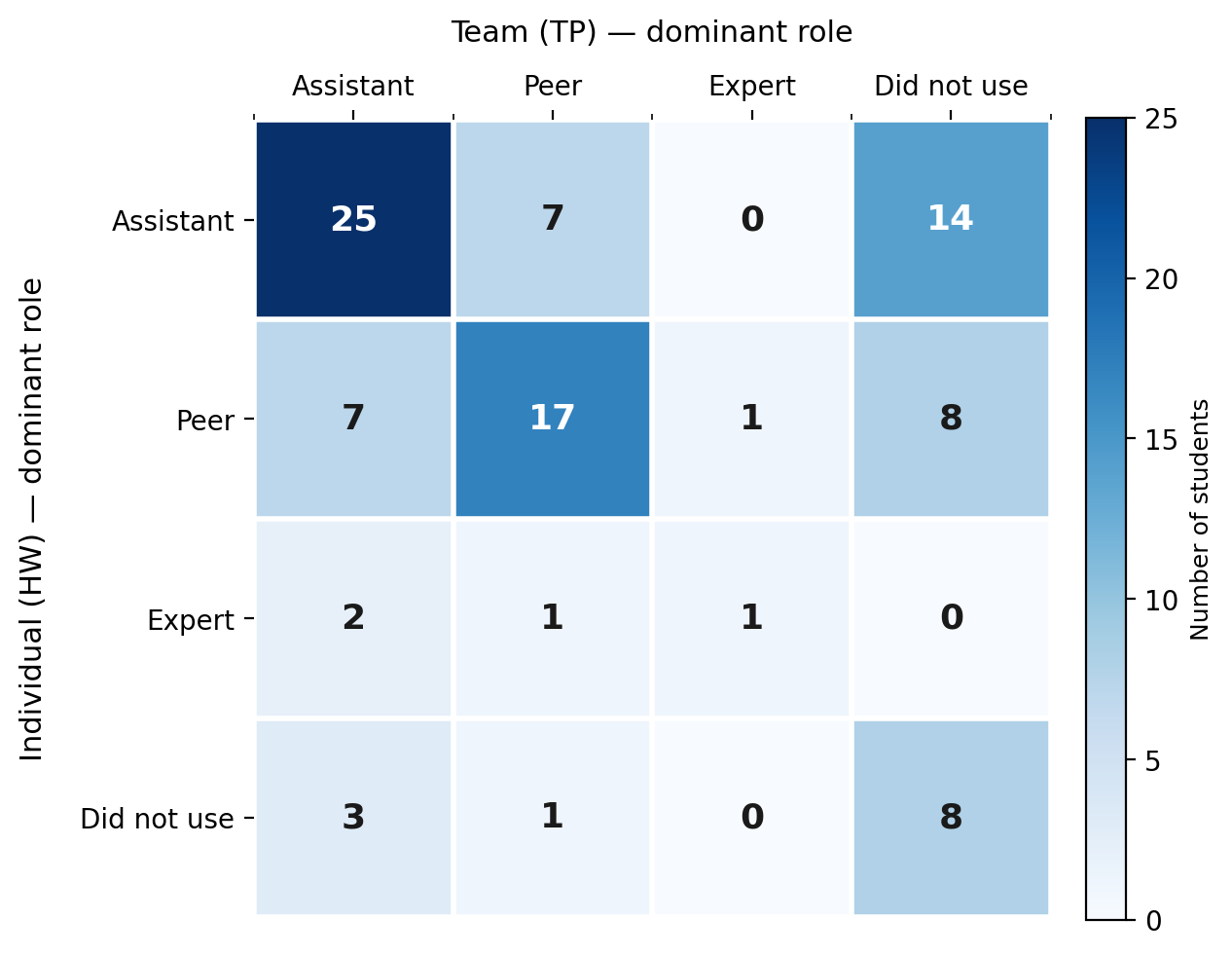}
  \caption{Within-student transitions in dominant LLM role
    framing from individual (HW) to team (TP) assignments
    ($n = 95$). Cell color intensity reflects the number of
    students making each transition, with darker blue
    indicating larger flows. The most common transitions were
    stability within the Assistant role (25 students) and
    within the Peer role (17 students), while 14 students
    moved from Assistant to non-use and 8 moved from Peer to
    non-use. Expert was rare as a dominant TP role (2 students
    total) reflecting its low prevalence in team contexts.}
  \Description{A 4 by 4 heatmap matrix showing how students'
    dominant LLM role in individual assignments mapped to
    their dominant role in team assignments. Rows represent
    HW roles (Assistant, Peer, Expert, Did not use) and
    columns represent TP roles (Assistant, Peer, Expert, Did
    not use). Darker cells indicate more students making that
    transition. The darkest cell is Assistant to Assistant
    with 25 students. Expert appears in both rows and columns
    but with very small counts.}
  \label{fig:transition}
\end{figure}

\subsection{RQ4: Within-Student Transition Analysis}

Aggregate percentages, while informative, can obscure whether
observed shifts reflect genuine behavioral change at the
individual level or merely compositional differences in who
responded to which surveys. The within-student transition
analysis addresses this directly.

Each student was classified as an LLM user or non-user based
on their dominant pattern across HW assignments and separately
across TP assignments. One student of the 96 consented
participants had no analyzable TP responses and could not be
assigned a dominant TP pattern; they are excluded from this
analysis only, yielding an analytic sample of 95. Of these
95 students, 67 (70.5\%) used LLMs predominantly in both
contexts. However, 18 students (18.9\%) who consistently used
LLMs across individual assignments stopped using them when
working in teams. Only 3 students (3.2\%) moved in the
opposite direction. Seven students (7.4\%) did not use LLMs
in either context. A Wilcoxon signed-rank test confirmed that
this asymmetric within-student shift was statistically
significant ($W = 254.0$, $p < .0001$), with students using
LLMs in a mean of 2.52 out of 3 individual assignments versus
1.93 out of 3 team assignments.

The role framing transition matrix further reveals how
individual students' dominant framing shifted across contexts.
Among students who primarily framed the LLM as an Assistant
in HW, 25 maintained that framing in TP while 14 shifted to
non-use. Among students who primarily framed the LLM as a
Peer in HW, 17 maintained Peer framing in TP while 8 shifted
to non-use. These transitions suggest that the behavioral
shift was not driven by a single student subgroup but
represented a broad movement across role types toward
disengagement when work became collaborative.

\section{Discussion}

The results of this study present a picture that is more
nuanced and more theoretically interesting than a simple
suppression narrative. LLM use did not merely decline when
students began working in teams---it collapsed at the moment
of team formation, then largely recovered as teams matured.
Understanding why requires engaging with what the data reveal
about the mechanisms underlying that arc.

\subsection{The Transition as the Critical Inflection Point}

The trajectory of LLM use across the six assignments reflects
the combined influence of two factors: task type and social
context. TP1 was a planning and scoping task with no coding
component, which naturally reduced opportunities for LLM use
relative to the coding-heavy individual homeworks. Task
structure is therefore a meaningful predictor of LLM
engagement in its own right, and course designers should
anticipate that non-coding tasks may suppress LLM use in
ways unrelated to social context.

Yet task type alone does not account for the full pattern.
TP2 and TP3 both required substantive coding and data
analysis comparable in technical demand to HW4 and HW5, yet
usage remained lower (75.0\% and 78.1\%) than the individual
homework average of 84\%, and verification rates never
recovered to individual assignment levels. This persistent
gap on technically equivalent tasks is consistent with
collaborative context being associated with reduced deliberate
LLM engagement, a pattern that aligns with what collaborative
learning research would predict about diffusion of
responsibility and unsettled accountability norms in team
settings~\cite{stahl2006}. As Penney et al.\ observe,
students gravitate toward LLMs precisely for their
low-social-pressure character~\cite{penney2025}; when the
team context itself generates social pressure, that calculus
may invert. The recovery at TP2 and TP3 suggests that as
teams stabilized and norms settled, students reintegrated
their individual AI workflows into collaborative
practice, but verification practices, once lost, did not
return.

\subsection{What Persists After Recovery: The Verification Gap}

While usage frequency and prompting volume largely recovered
by TP2 and TP3, the verification data tell a more troubling
story. Test-based verification dropped from 33.2\% on
individual assignments to 13.8\% across team assignments
overall, and never fully recovered. This shift carries the
largest effect size among the categorical measures
($V = 0.357$), and we consider it the most practically
significant finding given the direct connection between
test-based verification and learning outcomes documented
in prior work~\cite{poitras2024}.

To ensure this finding is not simply an artifact of reduced
LLM use at TP1 (where 54 of 92 respondents did not use the
LLM at all), we recomputed verification rates restricted to
student-assignment instances where the LLM was actually used.
The shift remains significant ($\chi^2(3) = 26.01$,
$p < .0001$, $V = 0.205$, small effect). Among users only,
non-verification rose from 3.7\% to 8.1\%, and test-based
verification among users fell from 59.5\% to 20.5\%, a
larger drop than the unconditional figure. At TP1 specifically,
among the 38 students who used the LLM, only 13.2\% ran tests
and 21.1\% reported no verification at all. The decline in
rigorous verification is not simply an artifact of reduced use.

The particular fragility of test-based verification in team
contexts is worth dwelling on. Manual reasoning (walking
through logic, checking steps) remained stable across
contexts (42.4\% vs.\ 42.2\%). What disappeared was the more
demanding practice of actually running code against generated
outputs. This distinction matters because running tests is the
verification behavior most directly tied to catching errors
in AI-generated code~\cite{liu2023evalplus}
and to building the understanding that distinguishes
productive from passive AI use~\cite{poitras2024}. Its
collapse in team settings, even after usage and prompting
recovered, suggests that collaborative contexts may be
associated with a persistent degradation in the quality of
AI interaction that outlasts the initial disruption of team
formation.

One plausible mechanism is diffusion of responsibility: in a
team, the implicit assumption that someone else will catch
errors may reduce each individual's motivation to verify
thoroughly, consistent with theoretical accounts of diffusion
of responsibility in collaborative work~\cite{stahl2006}.

\subsection{Role Framing and the Persistence of Engagement}

The within-student transition analysis offers a further
refinement of the aggregate picture. Among students who
continued using LLMs in team settings, the stability of
iterative refinement as a technique (27.5\% in HW vs.\
27.4\% in TP) suggests that students who remained engaged
did so with comparable deliberateness to their individual
assignment behavior. The behavioral shift was driven not by
a degradation in the quality of engagement among continuing
users, but by a substantial increase in the proportion of
students who disengaged entirely: 18 students (18.9\%) who
consistently used LLMs across individual assignments stopped
using them when working in teams, while only 3 moved in the
opposite direction.

This asymmetry has an important practical implication.
Efforts to improve AI use in team settings should focus less
on improving the quality of interaction among students who
are already engaging and more on understanding why a
significant minority disengage completely at the onset of
collaboration. Whether this disengagement reflects social
pressure~\cite{qu2026, adnin2025}, perceived redundancy
given division of labor, or genuine re-evaluation of AI's
utility in collaborative contexts is a question our data
cannot resolve but that future work should investigate
directly.

\subsection{Implications for Course and Assignment Design}

These findings suggest three concrete design implications for
computing educators.

First, both task type and the onset of team work are design
variables worth attending to. Non-coding tasks like TP1 may
suppress LLM use in ways unrelated to social context, and
instructors should consider designing team entry points that
include meaningful coding components if they wish to observe
and scaffold AI behavior from the outset of collaboration.
The transition into team work may also benefit from more
explicit scaffolding around AI use.
A targeted intervention at the start of the collaborative
phase, explicitly discussing how AI tools should be used
within teams, who is responsible for verification, and how
AI contributions should be attributed, could help address
the behavioral disruption observed at TP1 and the
verification degradation that persists beyond it.

Second, verification should be treated as an individually
accountable practice within team assignments. Assessment
rubrics that explicitly require evidence of output
verification, such as attached test logs, documented
debugging processes, or reflection on what was checked and
how, would make this practice visible and accountable in
ways that current team assignments do not.

Third, the interleaved design of this course provides a
natural scaffold that purely sequential designs lack. The
rapid recovery at TP2 suggests that returning to individual
work between team milestones may help students maintain their
AI practices across both contexts. Instructors designing
courses with extended team phases and no intervening
individual work may observe more persistent suppression than
we found here.

\subsection{Limitations}

Several limitations bound the interpretation of these
findings. The study was conducted across two sections of a
single upper-division course at one institution, which
restricts generalizability. The course serves students with
substantial prior programming experience; findings may differ
in introductory courses, in courses with different AI use
policies, or in disciplines outside computing. Transferability
to other course types, institutions, or student populations
should therefore be assumed cautiously, and replication
across diverse course contexts is an important direction for
future work. Additionally, we did not collect demographic
data on participating students, which limits our ability to
assess whether observed patterns vary by background,
experience level, or other individual characteristics.

All behavioral measures are self-reported, which introduces
the possibility of social desirability bias. Students may
have underreported LLM use in team settings if they perceived
it as less acceptable, or reported verification behaviors
more generously than their actual practice warranted. The
direction of this bias, if present, would tend to attenuate
the differences we observed, suggesting the true behavioral
shift may be larger than our data indicate. Future work
could complement self-report surveys with behavioral trace
data (such as LLM interaction logs or API usage records)
to reduce reliance on self-report and enable more
objective verification of student behavior across assignment
contexts.

Survey participation was voluntary, which introduces the
possibility of systematic non-response bias. Students who
chose not to complete a given survey may differ meaningfully
from those who did, for example in their LLM usage habits
or attitudes toward AI disclosure. Response counts ranged
from 90 to 96 across the six assignments. Students who
missed one or more surveys were not excluded from the
analysis; each student contributes data for the assignments
on which they responded. We did not impute missing values.
Because non-response was not systematically tracked against
student characteristics, we cannot rule out the possibility
that the observed patterns reflect in part who chose to
respond rather than the full population of consenting
participants.

Finally, while the within-student design controls for
individual differences, it cannot establish causality. The
observed behavioral shifts are associated with the transition
to team-based work but may reflect unmeasured features of
the specific assignments, the team formation process, or
other course-level factors. In particular, TP1 was a
non-coding planning task, which confounds social context
with task type for that specific data point. We address this
by examining TP2 and TP3 separately, both of which involved
substantive coding comparable to individual homework
assignments, and note that the verification gap persists
across all three team milestones. Claims about the
independent effect of collaborative context on LLM behavior
should therefore be interpreted with appropriate caution
given this partial confound.

\section{Conclusion}

This paper examined how students' use of LLMs changes as they
move from individual work to team-based assignments. Across a
semester-long study, we observed that this transition is not
seamless. LLM use declines sharply at the onset of
collaboration, partially recovers as teams stabilize, and
remains different in important ways from individual work.

In particular, while overall usage and prompting behavior
rebound over time, verification practices do not. Students
were consistently less likely to run tests or systematically
check AI-generated outputs in team settings, even when
working on comparable coding tasks. This suggests that
collaborative contexts may reshape not only whether students
use LLMs, but how carefully they engage with them.

These findings highlight the transition into team work as a
moment worth attending to in computing education. Course
designs may not always account for the implications of this
shift for how students interact with AI tools. In particular,
practices such as output verification, central to effective
AI-assisted learning, may require more explicit support in
collaborative settings.

More broadly, this work underscores that LLM use is not a
fixed individual behavior. It is shaped by context, task
structure, and social dynamics. Understanding how these
factors interact will be important for designing learning
environments that support both productive and responsible use
of generative AI.

\bibliographystyle{unsrt}
\bibliography{references}

\end{document}